\documentclass[%
 aip,
cp,  % Conference Proceedings
 amsmath,amssymb,%nobibnotes,
 reprint,%
]{revtex4-2}

\usepackage{graphicx}% Include figure files
\usepackage{dcolumn}% Align table columns on decimal point
\usepackage{bm}% bold math
\usepackage{mhchem}
\usepackage[utf8]{inputenc}
\usepackage[T1]{fontenc}
\usepackage{mathptmx} 

\begin{document}

\title{Production of GEM-like structures using laser-cutting techniques}% Force line breaks with \\

\author{L.\,M. Ramos} % Write as First name Surname
 \email[Corresponding author: ]{leticia.m.ramos@email.ucr.edu}
\affiliation{Department of Physics and Astronomy, University of California Riverside, Riverside, California 92507, USA}

\author{A.\,F.\,V. Cortez}
 \email{acortez@camk.edu.pl}
\affiliation{Astrocent, Nicolaus Copernicus Astronomical Center of the Polish Academy of Sciences, Rektorska 4, 00-614 Warsaw, Poland}%

\author{M.~Ku\'zniak}
 \email{mkuzniak@camk.edu.pl}
\affiliation{Astrocent, Nicolaus Copernicus Astronomical Center of the Polish Academy of Sciences, Rektorska 4, 00-614 Warsaw, Poland}%

\author{A.~Gnat}
\affiliation{%
Warsaw University of Technology
}%
\author{M.~Ku\'zwa}
\affiliation{Astrocent, Nicolaus Copernicus Astronomical Center of the Polish Academy of Sciences, Rektorska 4, 00-614 Warsaw, Poland}
\author{G.~Nieradka}
\affiliation{Astrocent, Nicolaus Copernicus Astronomical Center of the Polish Academy of Sciences, Rektorska 4, 00-614 Warsaw, Poland}
\author{T.~Sworobowicz}
\affiliation{Astrocent, Nicolaus Copernicus Astronomical Center of the Polish Academy of Sciences, Rektorska 4, 00-614 Warsaw, Poland}%
\author{S.~Westerdale}
 \email{shawn.westerdale@ucr.edu}
\affiliation{Department of Physics and Astronomy, University of California Riverside, Riverside, California 92507, USA}

\date{\today} % It is always \today, today, but any date may be explicitly specified
              % Not printed for conference proceedings

\begin{abstract}
It has previously been proposed to enhance the ionization yield of liquid argon time projection chambers (LArTPC) used in dark matter and neutrino experiments by loading LAr with dopants with ionization energies below the energy of LAr scintillation photons. 
While dual-phase LArTPCs have excellent sensitivity to single ionization electrons, granting some sub-keV thresholds, their compatibility with photosensitive dopants is hindered by gas-phase electroluminescence photons re-ionizing the dopants, creating a positive feedback loop.
This challenge can be addressed by optically decoupling the gaseous and liquid phases with a barrier that transmits electrons.
One potential way to do this uses a pair of structures based on Gaseous Electron Multipliers (GEMs) with mis-aligned holes.
Rather than amplifying electron signals in gas pockets within their holes, their holes will be filled with LAr and a lower bias voltage will be applied, so that incident drifting electrons are drawn into the holes but not amplified. Instead, amplification will occur in the gas phase above the structures, as typical for dual-phase TPCs.
% GEM-like charge amplification structures, including FAT-GEMs, can find applications in TPCs for dark matter, neutrino and generally rare event search experiments.
% A new concept of GEM-like structures was recently proposed. In this concept a double-stack of GEM-like structures is used to optically decouple LAr and GAr regions from a dual-phase TPC. 
Its core element is a GEM-like structure machined from polyethylene naphthalate (PEN). 
Since PEN scintillates in the visible spectrum, the risk of increased radioactivity due to the larger mass compared to traditional wire grids is negated by the potential to veto its own radioactivity. As such, these structures may also be a useful alternative to wire grids.
% Due to its intrinsic characteristics, such structure presents several advantages. On the one hand, it will work as self-veto regarding its own background, on the other hand it opens the possibility for doping LAr with dopants with very low ionisation energies (solving one of the main technical challenges related with optical positive feedback by having two layers of mismatched holes) while enabling at the same time the scaling up of such detectors.
In this work, we report the newest developments on the production of GEM-like structures using laser-based techniques, namely the manufacture of the first batch PEN and PMMA-based GEM-like structures. This process allows low-cost, reproducible fabrication of a high volume of such structures. In addition to being a low radioactive technique, we expect that it will allow the scaling up of the production of these structures at a reduced cost. First tests indicate good electrical stability, while the performance assessment is still ongoing.
\end{abstract}

\maketitle

\section{\label{sec:intro}Introduction}

WIMPs can be directly detected by dual-phase Ar-based TPCs, such as those developed by the DarkSide collaboration~\cite{agnes2023}. Future detectors will search in the lower energy region and focus on increasing the sensitivity, by relying on S2 signals and by improving S2 amplification. 
A possible improvement, currently being studied at the University of California, Riverside is the use of dopants with very low ionization energies, which may enhance the medium's ionization yield.
``Photosensitive dopants'' with ionization energies below the energy of scintillation photons can also convert S1 into additional S2. 
However, S2 photons may leak back into the LAr and cause additional ionization, resulting in a positive feedback loop.
To prevent this feedback, one possibility is to construct a structure similar to a gaseous electron multiplier (GEM)~\cite{sauli2016} that transmits electrons but blocks VUV photons. This can be done with two stacked GEMs with offset patterns, operated fully in the liquid phase, beneath a gas pocket in traditional dual-phase LArTPCs, with applied voltage too small to induce electroluminescence. To self-veto radiogenic backgrounds from the GEMs, 125~$\mu$m-thick polyethylene naphthalate (PEN), which scintillates, was used. 
The first batch used a hexagonal hole pattern, which optimizes surface coverage (higher hole density and hence better transparency) with 200~$\mu$m-diameter holes, and 1000~$\mu$m pitch.
% For the production of the first batch, a hexagonal hole-pattern was selected, as it provides a better surface coverage (higher hole density and as consequence higher open area) aiming at a 200~$\mu$m diameter hole, and a 1000~$\mu$m hole-pitch. 
A PEDOT:PSS (Clevios\textsuperscript{\texttrademark}) transparent conductive coating produced the electrodes, improving light collection of the readout plane while blocking VUV photons that may induce feedback.
% For the electrode material, PEDOT (Clevios) was chosen, due to the possibility of producing electrodes with different optical transmissions, that will allow to on the one hand improve the light collection on the readout plane, but at the same time minimize positive optical feedback by making the last electrode opaque to VUV scintillation. 
To improve the gain and minimize re-ionization, two such structures can be used. By shifting the holes' positions (misalignment), these structures prevent electroluminescence photons from the gas pocket above them from entering the LAr. Nevertheless, cost-effective, radiopure, and scalable production of these solutions poses additional challenges. 
%Since the invention of the first GEM by Fabio Sauli at CERN in 1997, GEM-like structures have been gaining relevance in several areas ranging from physics to medicine. Typically, these structures consist of a thin, metal-clad polymer foil, chemically etched with a high density of holes (typically 10 to 100 per mm$^2$). By applying a voltage difference between the different electrodes, a strong electric field is generated inside the holes that is capable of producing charge avalanches and/or secondary scintillation, leading to gains up to 10$^ 4$ (in a single GEM-like structure), positioning GEM-like structures as an interesting solution for both charge and light amplification. 

Despite recent advances observed in the development of MPGD technology, in particular of GEM-like structures, the methods of fabricating GEMs are limited~\cite{sauli2016, wu2023}. GEM production can be divided into two main approaches: 1) micro-hole production via photolithography and chemical etching or 2) mechanical hole drilling ~\cite{wu2023}. While the former may exhibit problems due to displacement of the holes in the dielectric and electrode layers due to alignment inaccuracies, the latter may result in metal burrs during drilling, which may cause electronic discharge~\cite{wu2023}. 

An alternative approach was developed at the University of Tokyo~\cite{fujiwara2018}, using doped photosensitive etchings (e.g. PEG and PEG3C), and a process similar to the lift-off method in the metal deposition step. Since the hole drilling relies entirely on photolithography, this approach displays two major challenges: First, uneven doping leads to rough hole walls, which manifest as tip structures in the holes, resulting in a greater discharge probability. Second, etching processes that rely on the photosensitive properties of special doped materials (such as Ag) cannot be applied to other glass materials with lower radioactive background and high-dielectric-strength. In addition, processes such as lift-off will cause curling, difficult to eliminate, and which can increase the electrical discharge probability.

In this work, the production of PEN-based GEM structures using laser-cutting techniques is presented. A novel GEM fabrication method is proposed to solve the technical limitations of current approaches, namely in what concerns the producing of conical (or bi-conical) GEM-holes in thick structures, that will allow to improve the electron transparency while minimizing possible problems from charge-buildup in the GEM-like holes. Due to its characteristics, this technique is expected to allow the development of large area solutions, and easy production scaling, being adequate for the production requirements of dark matter experiments as an alternative radiopure technique.

\section{\label{sec:production}GEM Fabrication Process}
The production of GEMs was performed at Astrocent and Warsaw University of Technology (Poland)~\cite{kuzniak20212024, leardini2024}. For the production of these structures PEN sheets with different thicknesses were used (acquired from Teijin-DuPont under the brand name of Teonex, or from Goodfellow).

\begin{figure}[h]
    \centering
\includegraphics[width=0.3\linewidth]{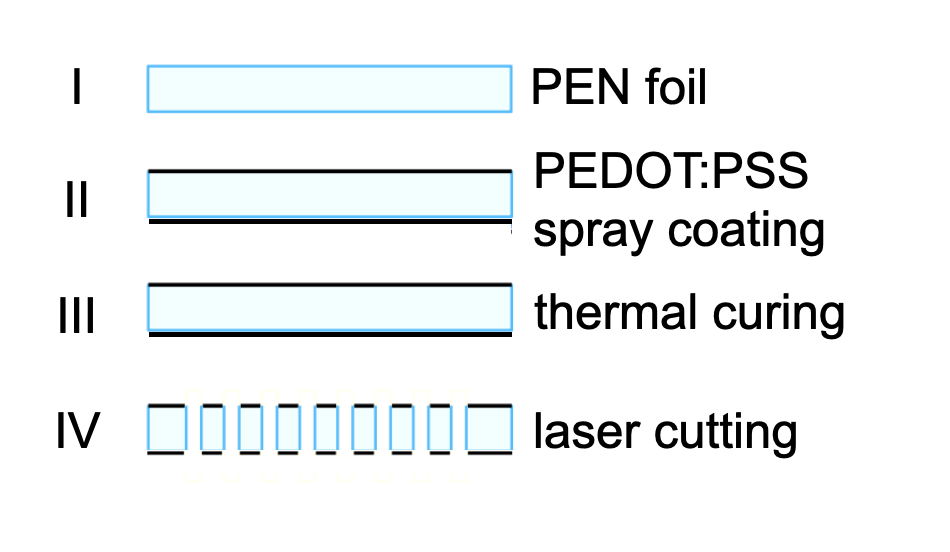}
    \caption{Steps of the fabrication sequence for a PEN-based GEM, adapted from \cite{kuzniak20212024}\label{fig:gem}}
\end{figure}

Fabrication steps are shown in Fig.~\ref{fig:gem}. The production of transparent conductive coatings was performed in an ISO7 cleanroom at CEZAMAT (Warsaw, Poland).
Before applying the coatings to the different PEN substrates, these were ultrasonically washed for 1~h in a solution of Alconox® in ultrapure rinse in UPW (ultrapure water) alone and drying. The edges of each tile were then masked with a plywood mask in order to ensure proper electrical insulation, minimizing the probability of electrical discharges through the edges of the tiles. After diluting PEDOT:PSS with a solution of isopropanol and UPW, coatings were deposited using a hand-held commercial airbrush (Paasche Talon TG-3F), supplied with a fan head permitting to uniformly deposit up to a 80~mm wide slit. The target thickness of the coatings, approximately 50~nm, was achieved by adjusting the amount of PEDOT:PSS in the solution relative to the area of the substrate. Deposition was performed in three passes for each substrate side.
Coatings were then cured in an oven at 80~C for 30~min. The coating displayed a surface resistivity of about the 30\,k$\Omega$/sq, at room temperature. After these initial tests the structures were stored in the dark, and taken just for the production of the hole patterns.

\begin{figure}[h]
    \centering
\includegraphics[width=0.3\linewidth]{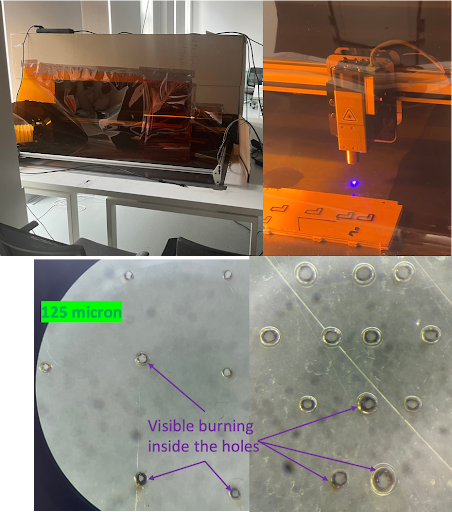}~~\includegraphics[width=0.27\linewidth]{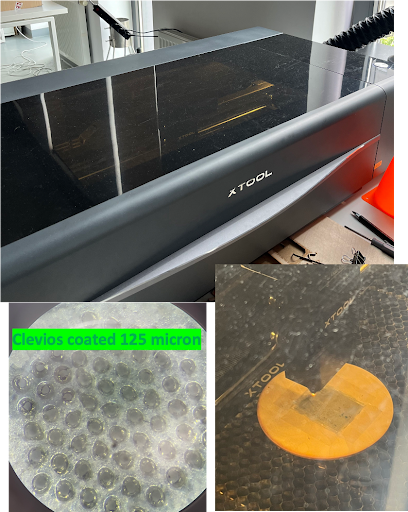}
    \caption{Blue light diode laser setup (top-left) and results of the holes obtained with this setup (bottom-left). Infrared CO$_2$ laser (top-right) and results and holes obtained with the setup (bottom-right).\label{fig:laser}}
\end{figure}

\paragraph{\bf Laser-cutting tests.}
For the testing of the laser-cutting technique and for the production of the first batch of GEMs, samples with three different thicknesses were used: 25, 50 and 125~$\mu$m-thick PEN sheets. All PEN samples used in this initial test were thin-coated with PEDOT:PSS, displaying similar surface resistance values. 
Two setups available at the Warsaw University of Technology were tested. A summary of the main characteristics of these systems follows.
\begin{itemize}
    \item \textbf{Blue light diode laser}\\
    ARCO~1010 setup with OptLasers blue diode laser with a 6~W power and 445~nm, was used.
    This setup can control the laser power, duration of the pulse, and number of repetitions. To perform these initial studies, a Python script was developed to sequentially produce holes with different configurations and number of repetitions. Later, the settings were evaluated by visual inspection and the dimension and variation on the hole size were determined optically with the help of specialized software. Figure~\ref{fig:laser} shows the setup used (top-left) and holes produced by the laser setup in  a 125~$\mu$m-thick PEN foil (bottom-left). The range of parameters studied in this setup can be found in Table~1. Using this setup, it was only possible to produce holes with minimal visible burn marks for \,25 $\mu$m-thick PEN foils was possible to produce holes with minimal visible burning marks, while for thicker foils there was no configuration that could result in clean holes (50 $\mu$m or 125~$\mu$m-thick PEN).
    
    \item \textbf{Infrared CO$_2$ laser}\\
    This cutting tests used the xTool~P2 with a 55~W power CO$_2$ laser. This setup can control the laser power, duration of the pulse (via the speed), diameter or the beam and the number of repetitions. To perform these initial studies (sequentially produce holes with the different configurations and number of repetitions) the proprietary software from xTool was used. A routine was developed to produce holes with different configurations and numbers of repetitions. No burning was visible on the bare 50~$\mu$m and 125~$\mu$m-thick PEN, even when using 1$\%$ of the laser power was found to be sufficient. Using the current setup, the smallest hole we can consistently cut is around 200~microns, but it is possible to improve by adjusting the aperture of the laser (beam size). Another important conclusion of this work was that the speed did not affect the hole size.
\end{itemize}

\begin{table}[h]
\centering
 \begin{tabular}{c||c|c|c|c|c} 
 Laser & Power (\%) & Duration (sec) & Repetitions & Hole diameter (mm) & $\sigma$ \\ [0.5ex] 
 \hline
 Blue &  1\% - 20\% & 0.1 – 2 sec & 1 – 20  &  - & -\\ 
 Infrared & 1 - 10$\%$ & 25 – 200 mm/sec & 1  &  0.2 mm & 0.026 mm\\ [1ex] 
 \end{tabular}
 \caption{Summary of laser parameters used in this work and minimum hole diameter achieved in a 125~$\mu$m-thick PEN foil.}
\label{tab}
\end{table}

\begin{figure}[h!]
    \centering
\includegraphics[width=0.18\linewidth]{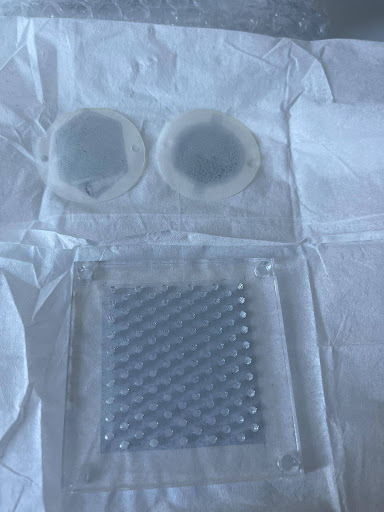}~~~\includegraphics[width=0.18\linewidth]{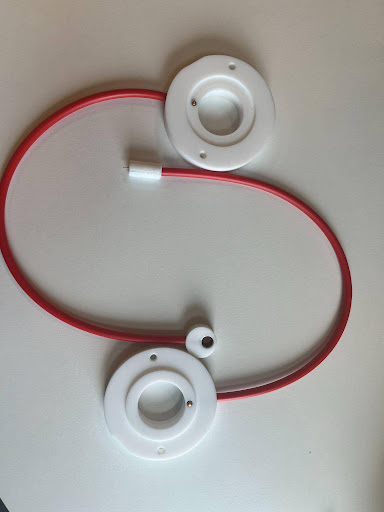}~~~\includegraphics[width=0.18\linewidth]{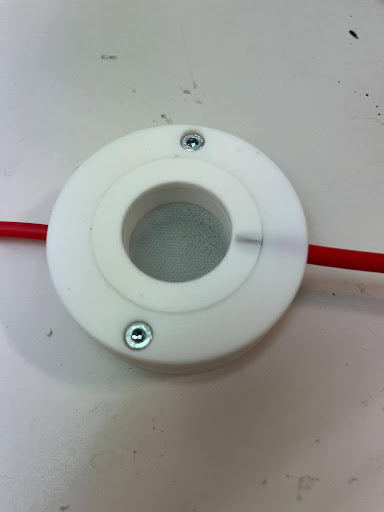}
    \caption{Examples of the PEN GEM and PMMA-based FAT-GEM structures produced with the CO$_2$ laser (left), setup used for the testing of the electrical stability of the round structures (center) and assembled with one of the structures (right).}
    \label{fig:uv}
\end{figure}

\paragraph{\bf PEN-based structures} Following the production of the hole-pattern in these structures, using the laser configuration that provided the best results, electrical stability tests were conducted on the PEN-based GEM structures. The objective of these tests was to assess if the integrity of the structures were affected by the laser-cutting process, as presence of conductive residue inside the holes could short the electrodes or increase the probability of electrical discharges. To conduct these tests, a simple setup, developed at Astrocent, was used. The setup used can be seen in Fig.~\ref{fig:uv}. By applying a voltage between the two electrodes and monitoring the current, it is possible to understand the impact of laser-cutting techniques on the electrical insulation of these structures. All samples produced were successfully tested up to 800~V, without observing a significant increase in the drained current (from the power supply). In one of the structures, we tried to identify the breakdown voltage. In this case, we manage to reach a bias voltage of 7~kV, however we experienced significant discharges, that resulted in the damage of the structure.

\paragraph{\bf PMMA-based structures.} As a result of the success of these initial tests, the laser was used to produce FAT-GEMs (Field Assisted Transparent Gaseous Electroluminescence Multipliers)~\cite{kuzniak20212024,leardini2024}. The FAT-GEM consists of a 5\,mm-thick PMMA structure coated with a conductive material on both sides and, optionally, with wavelength shifting coating (TPB) inside the holes. As in the previous test with PEN foils, it was possible to produce the hole-pattern using the infrared CO$_2$ laser, without any burn marks. A sample structure of 5~cm $\times$ 5~cm, with a hexagonal hole-pattern with 2~mm-diameter holes and 4~mm pitch was produced. The holes displayed an uncertainty in the diameter of about~10\%, presenting a conical shape instead of the traditional cylindrical obtained by mechanical drilling, that resulted from the fact that the focal point of the laser was coincident with one of the PMMA surfaces. The structures produced passed the electrical stability/insulation requirements. The main concern during the development was the residual stress in PMMA introduced by laser drilling, clearly visible with crossed polarizers, and potentially jeopardizing the mechanical integrity in cryogenic conditions. The issue was tackled by an additional thermal annealing step after the hole-pattern productionIn order to confirm acceptable stress level post-annealing, the PMMA-based structure was immersed in liquid N$_2$, with no damage observed.

\section{\label{sec:conclusions}Conclusions}
In this work, PEN-based GEM and PMMA-based FAT-GEM structures were produced using laser-cutting techniques for the first time. For the application in a self-vetoing optical decoupler between gas and liquid in TPCs, the technique was fine-tuned to avoid the conductive residue from charred PEN. In addition to this, more generally, this novel fabrication method is expected to solve the technical limitations of currently used approaches, namely in what concerns producing conical/bi-conical holes in thicker structures, that will allow to improve the gain, while minimizing possible problems arising from charge-buildup in the holes. Due to its characteristics, this technique will allow the development of large area solutions, and easy production scaling at a lower cost, adequate for the requirements of dark matter experiments as an alternative clean and radiopure technique.

\begin{acknowledgments}
This project has received funding from the Polish National Science Centre (2022/47/B/ST2/02015), from the European Union’s research and innovation programmes: Horizon 2020 under grant agreement No 952480, and Horizon Europe under the Marie Skłodowska-Curie grant agreement No 101154972, from the International Research Agenda Programme AstroCeNT (MAB/2018/7), funded by the Foundation for Polish Science from the European Regional Development Fund, and from the U.S. Department of Energy grant No. DE-SC0025540.
Kind support from Grzegorz Pastuszak and Zygmunt Skiba (Warsaw University of Technology), as well as CEZAMAT cleanroom laboratory staff, is gratefully acknowledged.
\end{acknowledgments}

\nocite{*}
\bibliography{aipsamp}% Produces the bibliography via BibTeX.

\end{document}